\documentstyle[twoside,fleqn,espcrc2,epsfig]{article}

\newcommand{\AmS}{{\protect\the\textfont2
  A\kern-.1667em\lower.5ex\hbox{M}\kern-.125emS}}

\newcommand{\be}{\begin{equation}}
\newcommand{\ee}{\end{equation}}
\newcommand{\ba}{\begin{eqnarray}}
\newcommand{\ea}{\end{eqnarray}}

\title{
\vspace{-10.0mm}
\begin{flushright}
\small
HUB-EP-97/63
\end{flushright}
        Critical exponents in U(1) lattice gauge theory with a monopole term 
        \thanks{Contribution to Lattice '97, International Symposium, 
                Edinburgh, UK, 1997.
                This research was supported in part under DFG grant 
                Ke 250/13-1.}}

\author{G. Damm \address{Fachbereich Physik, Universit\"at Marburg,
		       D-35032 Marburg, Germany },
W.~Kerler $^{\mbox{\scriptsize a,}}$\address{Institut f\"ur Physik, 
Humboldt-Universit\"at, D-10115 Berlin, Germany}}

\begin{document}

\thispagestyle{empty}

\begin{abstract}
We investigate critical properties of the phase transition
in the four-dimensional compact U(1) lattice gauge theory supplemented
by a monopole term for values of the monopole coupling $\lambda$ such
that the transition is of second order. It has been previously shown that 
at $\lambda= 0.9$ the critical exponent is already characteristic of a 
second-order transition and that it is different from the one 
of the Gaussian case. In the present study we perform a finite size 
analysis at $\lambda=1.1$ to get information wether the value of this exponent 
is universal.
\end{abstract}

\maketitle

\section{INTRODUCTION}

Our investigations are based on the Wilson action supplemented by a
monopole term \cite{bs85}
\be
S=\beta \sum_{\mu>\nu,x} (1-\cos \Theta_{\mu\nu,x})+
\lambda \sum_{\rho,x} |M_{\rho,x}|
\label{a}
\ee
where $M_{\rho,x}=\epsilon_{\rho\sigma\mu\nu}
(\bar{\Theta}_{\mu\nu,x+\sigma}-\bar{\Theta}_{\mu\nu,x}) /4\pi$
and the physical flux $\bar{\Theta}_{\mu\nu,x}\in [-\pi,\pi)$ is related
to the plaquette angle $\Theta_{\mu\nu,x}\in (-4\pi,4\pi)$ by
$\Theta_{\mu\nu,x}=\bar{\Theta}_{\mu\nu,x}+2\pi n_{\mu\nu,x}$ \cite{dt80}.
We use periodic boundary conditions.

Studies of energy distributions with a newly developed dynamical
algorithm have indicated that with increasing $\lambda$ the strength of
the first order transition decreases such that the transition ultimately
gets of second order \cite{krw94,krw95a}. 
Recently the behavior at larger $\lambda$ has been investigated in more
detail \cite{krw97}. The unexpected phenomenon has been revealed that the 
phase transition persists up to very large values of $\lambda$, where the 
transition moves to large negative $\beta$. Further it has been observed that 
(within errors) the monopole density becomes constant in the second-order 
region. By a finite size analysis it has been shown that at $\lambda= 0.9$ 
the critical exponent is already characteristic of a second-order transition. 
Moreover, with the result $\nu=0.446(5)$ obtained, it turned out to be
different from the exponent $\frac{1}{2}$ of the Gaussian case. 

Similar results have been recently reported \cite{jln96} for the Wilson 
action extended by a double charge term with coupling $\gamma$. There
the first order transition, which occurs at $\gamma=0$, weakens with 
increasing $\gamma$ until it becomes of second order at a tricritical
point. While for the usual periodic boundary conditions the second
order region starts at negative $\gamma$, for a spherelike lattice 
it occurs already at $\gamma=0$ \cite{jln96}. This could be due to 
inhomogeneities which tend to weaken the transition \cite{krw96}. 
The result obtained in Ref.~\cite{jln96} for $\gamma=-0.2, -0.5$ is 
$\nu=0.365(8)$ which is smaller than the value mentioned above.
Thus the two types of actions obviously lead to different results.

In the present study we continue the investigations with the action
(\ref{a}) because the addition of a monopole term is attractive in view
of the close relation of monopoles to the phase structure. The increase 
of the finite-size effects with $\lambda$ observed in \cite{krw97} has 
hinted at a possible increase of $\nu$. This suggests to check if for 
larger $\lambda$ the same exponent is obtained, which is done here by
performing a finite size analysis at $\lambda=1.1$.

\begin{figure}[ht]
\hspace*{-12mm}
\psfig{figure=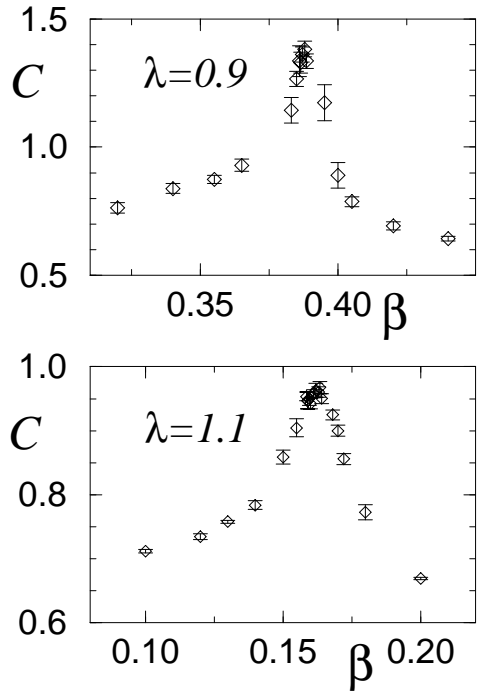,width=16cm,height=23cm}
\vspace*{-145mm}
\caption{Specific heat $C$ as function of $\beta$ for $\lambda=0.9$
and 1.1 on the $8^4$ lattice.}
\label{pic4}
\vspace*{-6mm}
\end{figure}

\section{DETERMINATION OF THE TRANSITION POINT}

In order to facilitate the determination of the transition point we have 
used the topological characterization of the phases \cite{krw94,krw95}.
It is based on the fact that there is an infinite network of monopole
current lines in the confining phase and no such network in the
Coulomb phase.  On finite lattices ``infinite'' is to be defined in
accordance with the boundary conditions \cite{krw96}. For the periodic
boundary conditions we are considering here, ``infinite'' is
equivalent to ``topologically nontrivial in all directions''.  While
for loops the topological characterization is straightforward, to
determine the topological properties of networks it is necessary to
perform a more elaborate analysis based on homotopy preserving
mappings \cite{krw94,krw95}.

Because the probability to find a network which is nontrivial in all 
directions takes values exactly 1 and 0 in the confining phase and in 
the Coulomb phase, respectively, it is a very advantageous order
parameter. Indeed, with an infinite system a single configuration would
be enough and on a finite lattice still few
configurations are sufficient to discriminate between the phases.

Since on finite lattices different order parameters lead to slightly 
different critical $\beta$, to determine the location of the maximum of 
the specific heat for larger $\lambda$
in an efficient way we first determine the critical $\beta$ from the
topological order parameter and then find the maximum of the specific
heat in an easy second step.

That finite size effects increase with $\lambda$ is indicated by the 
fact that the transition region becomes broader. From Figure 1 it can 
be seen that the width of the peak of the specific
heat increases and that its height decreases with increasing $\lambda$.

\section{CRITICAL BEHAVIOR}

In order to obtain information on the nature of the phase transition we 
have investigated the finite-size scaling behavior of the maximum of the 
specific heat $C_{\mbox{\scriptsize max}}$. It is expected to be
\be
C_{\mbox{\scriptsize max}} \sim L^d
\ee
if the phase transition is of first order and
\be
C_{\mbox{\scriptsize max}} \sim L^{\frac{\alpha}{\nu}}
\ee
if it is of second order, where $\alpha$ is the critical exponent of the
specific heat and $\nu$ the critical exponent of the correlation length.

In Figure 2 we present results of simulations for
$C_{\mbox{\scriptsize max}}$ on lattices with $L$ = 6, 8, 10, 12, for
$\lambda = 0.9$ obtained in Ref.~\cite{krw97} and for $\lambda = 1.1$ obtained 
here, at the respective values of $\beta_{\mbox{\scriptsize cr}}$.  The fit 
to these data gives the values for $\frac{\alpha}{\nu}$ shown in Table 1. They 
are clearly quite far from 4 and thus the transition is not of first order.

\begin{table}[h]
\caption{}
\begin{tabular*}{75mm}{@{}l@{\extracolsep{\fill}}lcc}
\hline 
 $\lambda$&0.9&1.1\\
\hline
 $\frac{\alpha}{\nu}$&0.485(35)&0.289(68)\\
\hline
 $\nu$&0.446(5)&0.466(8)\\
\hline
 $\beta_c(\infty)$&0.4059(5)& 0.1882 (34) \\
$a$&-1.99(6)& -2.29 (10)  \\
\hline
\end{tabular*}
\label{tab1}
\end{table}

\begin{figure}[ht]
\vspace*{14mm}
\hspace*{-6mm}
\psfig{figure=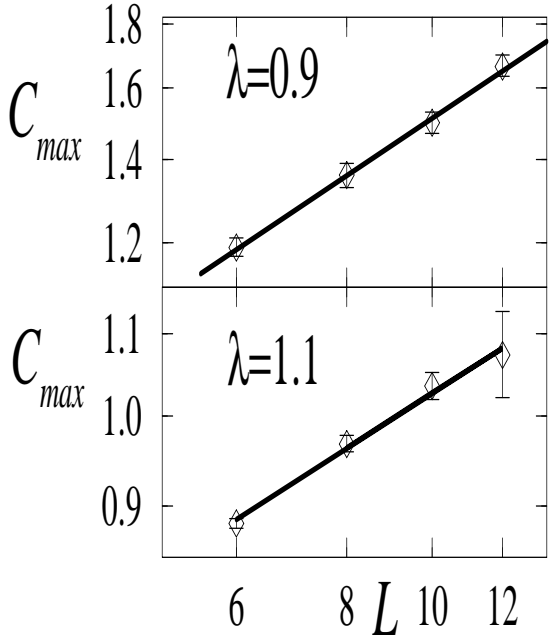,width=16cm,height=24.0cm}
\vspace*{-175mm}
\caption{Maximum of specific heat $C_{\mbox{\scriptsize max}}$ as function 
of lattice size $L$ for $\lambda=0.9$ and 1.1 at 
$\beta_{\mbox{\scriptsize cr}}$.}
\label{pic5}
\end{figure}

The results  for $\nu$ listed in Table 1 are obtained using the hyperscaling 
relation $\alpha = 2 - d\,\nu$. They are different from the value 
$\frac{1}{2}$ of the Gaussian case. There is a slight increase
with $\lambda$ which is beyond statistical errors.

The critical $\beta$ is expected to behave as
\be
\beta_{\mbox{\scriptsize cr}}(L)=\beta_{\mbox{\scriptsize cr}}(\infty) 
+a L^{-\frac{1}{\nu}}
\ee
 From this relation, using the values of $\nu$ in Table 1 and our data for
$\beta_{\mbox{\scriptsize cr}}(L)$, the numbers given for 
$\beta_{\mbox{\scriptsize cr}}(\infty)$ and $a$ in Table 1 are obtained.

\section{DISCUSSION}

The slight increase of $\nu$ with $\lambda$ observed could indicate a
nonuniversal behavior or even hint at the possibilty that $\nu$ ultimately 
reaches the Gaussian value. However, finite size analyses on larger lattices 
and for still larger $\lambda$ appear necessary to decide these questions.

Different values for $\nu$ here and in Ref.~\cite{jln96} could signal different
universality classes for the respective actions. One should, however,  also 
be aware of the possibility of obtaining different limits when approaching 
from different regions of coupling space \cite{c81}. Further the use of
different operators can lead to different values. Indeed, in an
investigation of gauge balls with the double charge action for $\gamma=-0.2$ 
the values $\nu \simeq 0.35$ and 0.5 have been reported \cite{cfjlns96}. 

It is interesting that recently theoretical arguments
have been presented \cite{aes97} predicting the set of exponents 1/3, 5/11 
and 1/2, which within errors are in agreement with the values observed in the 
simulations in Refs.~\cite{krw97,jln96,cfjlns96} and here. 

With the non-Gaussian exponents continuum limits of four-dimensional pure 
U(1) gauge theory, nontrivial and not asymptotically free, appears possible.
They would be different from the usual expectations for QED and would
not be seen in perturbative approaches. Possibly then monopoles survive in
the limit.  

\section*{ACKNOWLEDGMENT}

\hspace{3mm}
One of us (W.K.) wishes to thank M.~M\"uller-Preussker and his group for 
their kind hospitality.

\end{document}